# RELATIONAL PARADIGM IN ASKS AND ANSWERS

M. I. Suslova
Moscow society of naturalists, Russian philosophical society
A. A. Sidorova-Biryukova
Physical Faculty, M.V. Lomonosov Moscow State University
asidorova@mail.ru

SUMMARY
What is the reason for a minor success of relational ideas in scientific and public society? Are they actually that weird as they seem at the first glance? Here we attempt to look with a fresh eye at the main relational principles (derivative nature of space-time, action at a distance, and Mach's principle), revise some questions that relationalists often hear from their opponents, both physicists and wide audience, and probably find new evidence in favor of the relational concept.

INTRODUCTION
Relational paradigm does not currently dominate in physics. Orthodox physicists consider its principles artificial, not quite scientific, and even scholastic. Nevertheless, there are others, who consciously choose the relational viewpoint as the most adequate platform to represent the universe. Last time their number is notably growing. There is no need to list here the names, it suffices only to mention that, in the opinion of theoretical physicist Professor Lee Smolin (Perimeter Institute, Canada), the revolution in the 20$^{th}$-century physics can be called relational; and a unified theory combining the general relativity and quantum theory should appear as a completion of the relational revolution [1]. This viewpoint is shared by Professor of Moscow State University Yury Vladimirov, who is firmly convinced that relationism is the future of physics [2]. If their anticipations come true, the relational paradigm will become a fundamental component of the worldview system developed by human civilization. Albeit, now we still have time to take a closer look at its main ideas and assess whether they are actually strange and what place they should take in our vision of the world order.

The main principles of relationalism, according to [2], are the following: (1) space and time are conceived to have not fundamental but derivative nature; (2) physical interactions are described within the model of action at a distance, not short-range action; and (3) local properties of objects are considered as an approximation of the global properties of the entire world around them (Mach's principle).

These three principles complement each other and embrace every structural level of the universe. All of them are far from being obvious and need further explanation. Let us consider each of the three in turn.

# 1. DERIVATIVE NATURE OF SPACE-TIME

"According to the relational approach, space and time are not independent entities, as postulated by Isaak Newton, but represent an abstraction from relations between material objects (or events in which they participate)" – that is how the first relational principle is explained in [2]. While this explanation may probably satisfy physicists, the other part of the audience will likely find it difficult to imagine how "abstraction" arises from "relations between material objects". It is no surprise that most of people, and still most of physicists, prefer to treat space and time as an absolute environment containing material bodies, a stage on which the play of the universe is performed. Indeed, such view is consistent with our intuitive perception. But this is only at first glance.

The question of space-time is more difficult than it may seem. Our impression that it is elementary is misleading. In fact, being humans, we are extremely complex systems ourselves, the same are the processes of our interaction with the world; so what seems simple and natural to us does not have to be such in physical sense, that is, represent the fundamental elements of the universe.

To begin with, the very concept of space-time is not well defined in physics. This is clearly seen with the example of general relativity (GR) theory. (The example and the very idea of "the space-time emergence levels" are borrowed from [3].) Classical GR considers a 4-dimensional space-time manifold, on which one defines metric and material fields. The GR equations express the relations between these fields, and hence, describe the dynamics of curved space-time. But what is space-time? First, it can be the metric field and the related quantities: time intervals, spatial lengths. However, these quantities have no physical meaning until they contain material objects; therefore, the second version of space-time includes the material fields. Finally, it can be the 4-dimensional manifold itself, which version is the most close to our intuitive idea of space-time as a set of points of the space-time continuum. Unfortunately, this definition is also less meaningful. The fact is that the GR equations must be invariant for any smooth transformation of the manifold, which means that the physical properties of the system cannot be defined as functions of the absolute position of a point in this manifold.

Thus, there is no absolute space-time, which could be taken as a background for the action of force fields. As Carlo Rovelli put it, "No more fields on space-time, just fields on fields" [4]. Note that the failure to clearly define space-time does not only contradict our intuitive perception but also produces a philosophical problem, since the very notion of existence implies being at a certain place in space for a certain period of time, but what if the ontological status of space and time is questionable?

In this situation, the relational approach offers a natural solution. Instead of absolute space-time, one can use relative characteristics – both spatial and temporal. It can be said that mathematically perfect clocks and lengths are replaced by physical clocks and rods, which measure the temporal and spatial relations between events. These are precisely those "relations between material objects" mentioned above. Long time ago Gottfried Leibniz suggested that space and time are just instruments for convenient registration of relations between objects and events in the universe. Remarkable is how modern physicist Brian Greene comments on this preposition: «Although Newton's view, supported by his experimentally successful three laws of motion, held sway for more than two hundred years, Leibniz's conception, further developed by the Austrian physicist Ernst Mach, is much closer to our current picture» [5].

Above we described only the first step in space-time "elimination". Further problems arise when we try to find its smallest element. Searching for a quant of space-time inevitably leads one to a thought on why geometry is considered to be more fundamental than physics. Couldn't it be vice versa – geometric notions are generalized properties of physical objects? The secondary nature of space-time geometry is often discussed by specialists who develop a theory of quantum gravity. After all the attempts to quantize gravitational field directly (by dividing it into ever small pieces in space-time) have failed, physicists had to admit that the quantum of space-time may have nongeometric nature.

The idea is not new; similarly, the properties of a molecule of water have nothing to do with those of water as a continuous medium. In past decades, an enormous variety of structures were suggested as possible candidates to comprise space-time at the quantum level (see, e.g., [6]). Among the earliest is the classical concept of spin networks proposed by Roger Penrose in 1960s [7] and an example of most recent ones is the model of causal sets recently developed by Lee Smolin and collaborators [8]. Most of these structures are combinatorial, i.e., discrete entities obeying algebraic laws. This discreteness corresponds to the fundamental property of nature to be discrete at the elementary level. The quantized degrees of freedom mean that the usual continuous space-time breaks up into non-geometric elements, from which it has to emerge in the limit of an infinitely large sum of them.

Continuous space-time disappears together with all the related concepts, such as direction in space or time, geometric laws, metric fields, etc. All of them should emerge as new properties from the collective interaction of a very large number of elementary components. Moreover, it turns out that many different sets of such properties are possible and it is important to find arguments in favor of particularly those observed in our universe.

The above relational description of space-time is only a brief sketch, which leaves much room for clarification. A few steps further are made in the impromptu discussion below.

**Question**: If space is an "abstraction", then how should one treat its three-dimensionality? Length, width, and height are quantitative characteristics of mate-rial objects. Abstraction is a thought form; it cannot have quantitative characteristics of a material object otherwise it is no longer an abstraction but a material object.

**Answer**: There is no contradiction here, all mathematical objects are abstractions characterized by quantitative parameters; and the term "thought form" is quite appropriate. Indeed, many philosophic schools of the past represented space and time as mental forms, for example, master of Indian yoga "sees the motion picture of the cosmos going backward and forward on the screen of his consciousness: he knows in this way that time and space are dimensional forms of thought, displaying cosmic motion pictures" [9, Vol.1, p.45]. As for the three-dimensionality of space (or the four-dimensionality of space-time), the correct space-time signature (1, -1, -1, -1) naturally stems from the theory of binary systems of complex relations, developed within the relational framework [2].

**Question:** It is claimed that space emerges "from the relations between material objects"; does this mean that a single material object cannot produce space?

**Answer**: It depends on what is meant by a material object. In case it is macroobject, then it consists of a huge number of microobjects, whose interrelations define the macroobject's property to have size and, therefore, produce the embracing 3D space for it. The huge ensemble of elementary objects and their relationships creates what we call space. In case we are talking about a single microobject, then – yes, indeed, one elementary object cannot create continuous space since a very large (actually infinite) number of discrete entities are required to form something continuous. However, this is a sort of speculative question, because no single elementary object can exist in view of the primary role of relations, which involve at least two partners. Generally, the relational paradigm dislikes the idea to study systems isolated from the world. Instead, it takes on board the Mach's principle, which states that the universe should be considered in its integrity.

**Question:** If space emerges "from relations between material objects", what are the particular processes and mechanisms that bring these relations about?

**Answer:** Processes imply occurring in space and time, so it makes no sense to talk about them at the elementary level. But we can say that objects exchange their properties, beyond space and time. We touch upon this issue once again in the next section.

2. PHYSICAL INTERACTIONS AS ACTIONS AT A DISTANCE
The second principle of the relational concept concerns interaction between physical objects. There are principally two opposite ways to think about interaction in physics: either it is transferred from point to point between objects (short-range action) or the objects interact directly wherever they are located in space (action at a distance). The idea of short-range action is quite evident, but the action-at-a-distance scenario is not so easily understood, because our experience does not give us examples of how bodies "interact with each other directly at a distance" [2], i.e., with no agent transferring between them. Sometimes action at a distance is considered archaic, but this way of thinking definitely fits to the relational concept. Let us look at the arguments.

Analyzing the short-range action attentively, one can notice that this idea conceals an element of deception. Indeed, it is commonly accepted that all types of matter, including fields and substances, are quantized, that is, exist in the form of separate elementary portions. There is no other way for these elements to interact except by means of direct action, without any intermediate agent, in view of the absence of such. Perhaps, space-time itself could be suggested as such an agent, like it is in the GR theory? However, in view of the first relational principle (Section 1), space-time is deprived of independent existence, which means that elementary interaction cannot be embodied in anything expressed in spatio-temporal terms. In this situation, the action at a distance seems to be a more straight and natural concept.

No doubt, it is a challenge to describe any specific action-at-a-distance mechanisms and develop appropriate mathematical tools. But whatever they are, they should account for the primary role of interaction between objects with its properties determining the objects with their properties, and not vice versa. An illustration from the everyday life is child balance swing, which stands still on a playground and wait for a suitable pair of children, precisely a pair, who will have the right weight, height, and other parameters determined by the construction of the swing. By the way, the primacy of interaction is in line with the principle of a minimal number of basic elements, since at least two of them are required for one interaction event.

(An attitude towards relations between objects as something more important than the objects themselves seems natural far beyond the scope of physics; for example, when studying social models, we pay prior attention to the types of relations between people: friendship, sympathy, cooperation, etc., while their numerous specific realizations by various partners are of minor significance.)

From the relational viewpoint, interaction is not expressible in spatio-temporal terms but exists as possibility, which is realized as soon as a suitable pair of objects is found, for example, a source and a receiver. It is interesting to remind of a provocative question that relationalists often hear from their opponents: if there is nothing propagating between a source and detector of a signal, then where is the energy of the signal after it has been emitted from the first but has not yet arrived at the second? In our opinion, the question is an ill-posed one, the very formulation of which is wrong since it is inappropriate to speak in spatio-temporal language about something that occurs beyond space and time. Even the word "where" is inapplicable here.

Note that this ideology is not quite uncommon to conventional quantum-field theorists: they use something of the sort when transfer calculations from the coordinate into momentum space, where they are essentially simplified. The idea to get rid of time instants and space functions has led Richard Feynman to his famous method of diagrams and propagators. He described this remarkable idea in his Nobel lecture: «Instead of wave functions we could talk about this; that if a source of a certain kind emits a particle, and a detector is there to receive it, we can give the amplitude that the source will emit and the detector receive. We do this without specifying the exact instant that the source emits or the exact instant that any detector receives, without trying to specify the state of anything at any particular time in between, but by just finding the amplitude for the complete experiment. And, then we could discuss how that amplitude would change if you had a scattering sample in between, as you rotated and changed angles, and so on, without really having any wave functions» [10].

There is one more question, which concerns also the first relational principle (about space-time).

**Question**: If the action-at-a-distance concept implies "nothing" propagating between bodies, then how does space emerge from "nothing"?

**Answer**: It is meant that action at a distance implies no propagation of anything in space and time. As was mentioned above, interaction is prior to the objects and exists not as a process in space and time but as a rule, or a law, which defines the properties of the interacting objects. The collective behavior of an infinitely large number of objects, which we observe at the macrolevel, is perceived as a process occurring in space-time.

3. MACH'S PRINCIPLE
Finally, the third principle of the relational concept states that the local properties of matter are defined by the global properties of the entire world around (a version of the Mach's principle). This statement usually meets less counteraction than any of the previous two. We are already accustomed to think that the universe is an integrate system with all its parts tightly connected and dependent on each other. This system should have global properties and be ruled by global laws. However, this most easy-to-understand principle is also the most provocative one since it makes us think about the essence of these global laws and their relation to the local ones observed on practice.

**Question:** Why do we think that studying ever smaller parts of the world we get global knowledge about the universe?

**Answer:** Mach himself acknowledged that our intuitive perception and everyday experience are possibly not the right point to start from. He wrote [11]: "Nature does not begin with elements, as we are obliged to begin with them". We perceive the properties of the world locally, here and now, and do not always realize that what we are studying is not an isolated system, but a part of the single universe. The idea that complex can be studied by dividing it into simple is in the core of modern science, which is based on analysis. The corresponding mathematical instrument is the Newtonian apparatus of integral-differential calculus. Let us recall the Brian Greene's words cited above that despite its 200-years domination, this method has a number of promising alternatives today. The main feature of alternative methods is that, instead of continuous geometric concepts, they rely on discrete quantities ruled by algebraic laws. In fact, it has long been known that the laws of Euclid and Lobachevsky geometries can be written as systems of algebraic equations (following from the condition of zero Cayley-Menger or Gram determinants). If physics took the path of using algebraic rather than integral-differential calculus, the relational paradigm could be dominant.

Currently, a number of theories are being developed based on algebraic schemes, such as the systems of relations theory, the direct interaction theory, and the binary systems of complex relations theory [12]. They are successfully applied to construct physics within the relational framework. A detailed description of the results can be found in [12]. One example related to the Mach's principle is given below.

It is conventional in physics that a physical system is characterized by a certain functional quantity (action, Hamiltonian, etc.), which includes two types of terms: some stand for the free motion of the system's elements and the others account for the interaction between them. It is largely implied that the free motion of bodies occurs in space-time, which is independent of them. From the relational viewpoint, only terms of the second type have right to exist, while the legality of the first ones is in question. However, if one decides to adhere to the Mach's principle strictly, he should take into account the interaction not only between pairs of the system's elements but also between each of them and the rest of the world. A remarkable result is that careful consideration of such interaction with the third partners (in terms of the action-at-a-distance theory) adds new terms to the action functional that are exactly equal to those usually ascribed to free motion. Thus, the free motion of an object is "a veiled result of its interaction with all particles of the world around" [12].

In recent years evidence has been growing to show that a radically new way of thinking is required as an alternative to the mechanistic method of Newtonian physics. A deep and extensive survey of the problem is given in "The Systems View of Life" [13] by Fritjof Capra and Pier Luisi. The systems view is an attempt to elaborate a unifying vision of physical, biological, social phenomena as various forms of life and to find the most fundamental laws and patterns that are common to all of them. Its key components include complexity theory, nonlinear dynamics, fractal geometry, autopoiesis theory, models of organization and network structures. The main conclusion drawn by the authors is that now it is time to recon-struct our relations with the world in such a manner that collective interests will guarantee individual well-being and not vice versa, as it was until now and has brought the world to the brink of ecological (and now, when these words are written, also epidemiological) disaster.

CONCLUSION

Some people believe that science should not only describe how the world works, but also find the reason why the description is the best one. The question why one physical picture is better than another is largely a matter of taste and faith. It is important that scientists do not lose the ability to communicate and hear each other so that not to repeat the history of Babylon tower. Arguing and criticizing, different paradigms improve each other; sharing ideas, they complement each other; anyway the interaction leads to the progress.

Another question is that even if humanity can unite efforts in striving for a single theory of everything, what is the guarantee that collective mind chooses a right path. There are possibly two reasons for optimism. First, we all belong to this world, with all its laws working inside us, within every atom. That is why we can hope that our intuition formed by the experience of numerous generations (and even prebiotic evolution) gives us correct guidelines [14]. And there is one more trustworthy criterion, which is practice. As physicist David Deutsch wrote in "The Structure of Reality" [15], "Planes keep flying", meaning that we are able to realize our plans. Note that the same argument was formulated centuries ago by Indian sages, as Vedanta Sutras say: "... if all reasoning were unfounded, the whole course of the practical human life would have to come to an end" [16].

ACKNOWLEDGEMENTS

The authors are grateful to Prof. Yu. S. Vladimirov for his concern to our work on making the relational ideas popular, and to A. V. Soloviev for useful discussions and comments made to improve the article.